\documentclass[aps,prd,amsfonts,amssymb, amsmath, showpacs, showkeys, a4paper, nofootinbib, superscriptaddress, 11pt, raggedbottom]{revtex4-2}

\usepackage{braket}
\usepackage{bm}
\usepackage{graphicx}
\usepackage{slashed}
\usepackage{subfigure}
\usepackage{mathrsfs}
\usepackage{xcolor}
\usepackage{mathtools}
\usepackage{appendix}
\usepackage[normalem]{ulem}

\usepackage[export]{adjustbox}
\usepackage[utf8]{inputenc}
\usepackage[T1]{fontenc}
\usepackage{etoolbox}
\usepackage[font=scriptsize]{caption}
\usepackage{revsymb4-1}
\usepackage{blindtext}
\usepackage{hyperref}
\hypersetup{
    colorlinks=true,
    linkcolor=blue,
    filecolor=blue,      
    urlcolor=blue,
    }
\hypersetup{
colorlinks=true,
 citecolor=blue,
}

\def\beq{\begin{equation}}
\def\eeq{\end{equation}}
\def\beqa{\begin{eqnarray}}
\def\eeqa{\end{eqnarray}}
\def\rmd{\mathrm{d}}
\def\rmi{\mathrm{i}}
\def\rme{\mathrm{e}}

\begin{document}

\title{\Large Dynamical Casimir effect with screened scalar fields}

\author{Ana Lucía Baez-Camargo}
\email{lucia.baez@univie.ac.at}
\affiliation{Faculty of Physics, University of Vienna, Boltzmanngasse 5, 1090 Vienna, Austria}

\author{Daniel Hartley}
\email{danielhartley0@yahoo.com.au}
\affiliation{Independent researcher, Vienna, Austria}

\author{Christian K\"{a}ding}
\email{christian.kaeding@tuwien.ac.at}
\affiliation{Technische Universit\"at Wien, Atominstitut, Stadionallee 2, 1020 Vienna, Austria}

\author{Ivette Fuentes-Guridi}
\email{I.Fuentes-Guridi@soton.ac.uk}
\affiliation{School of Physics and Astronomy, University of Southampton, Southampton SO17 1BJ, United Kingdom}
\affiliation{Keble College, University of Oxford, Oxford OX1 3PG, United Kingdom}

\begin{abstract}
Understanding the nature of dark energy and dark matter is one of modern physics' greatest open problems. Scalar-tensor theories with screened scalar fields like the chameleon model are among the most popular proposed solutions. In this article, we present the first analysis of the impact of a chameleon field on the dynamical Casimir effect, whose main feature is the particle production associated with a resonant condition of boundary periodic motion in cavities. For this, we employ a recently developed method to compute the evolution of confined quantum scalar fields in a globally hyperbolic spacetime by means of time-dependent
Bogoliubov transformations. As a result, we show that particle production is reduced due to the presence of the chameleon field. In addition, our results for the Bogoliubov coefficients and the mean number of created particles agree with known results in the absence of a chameleon field. Our results initiate the discussion of the evolution of quantum fields on screened scalar field backgrounds.     

\end{abstract}

\keywords{Chameleons, Dynamical Casimir effect, Quantum field theory in curved spacetime}

\maketitle

% =========================================================================================================================================

\section{Introduction}\label{Intro}

Quantum field theory in curved spacetime (QFTCS) studies the behaviour of quantum fields propagating in a classical relativistic background geometry \cite{BD1984, wald1994quantum, parker_toms_2009}. This theory has predicted many physical phenomena, such as cosmological particle creation \cite{Parker1968, Parker1969, parker_toms_2009}, Hawking radiation \cite{Hawking1975} and the Unruh effect \cite{Unruh1976}, as well as the dynamical Casimir effect (DCE) \cite{moore1970quantum}, which refers to the generation of particles due to the motion of boundaries (see Refs.~\cite{dodonov2009dynamical, juarez2022short, dodonov2020fifty} for reviews). The first computations of the DCE were done in flat spacetime \cite{moore1970quantum, fulling1976radiation}. Over the past five decades numerous developments have appeared including distinct geometries of the cavities \cite{dalvit2006dynamical, pascoal2008dynamical, naylor2012towards}, entanglement generation \cite{friis2013entanglement, busch2014quantum, felicetti2014dynamical, romualdo2019entanglement}, and extensions to a few other metrics \cite{celeri2009action, lock2017dynamical}. Performing a mathematically rigorous study of the DCE is challenging due to the complexity of studying quantum field theory with dynamical boundary conditions \cite{juarez2023quantum, juarez2021quantum}. For this reason, within the framework of QFTCS, in Refs.~\cite{barbado2020,barbado2021}, some of the authors of the present work introduced a general method to compute the evolution of a confined quantum scalar field in a globally hyperbolic spacetime by means of a time-dependent Bogoliubov transformation. Part I \cite{barbado2020} considers spacetimes without boundaries or with timelike boundaries that remain static in some synchronous frame, while Part II \cite{barbado2021} considers spacetimes with timelike boundaries that do not remain static in any synchronous frame. 

QFTCS builds on the framework of general relativity (GR), which has proven to be a remarkably successful theory of gravity and cosmology \cite{wald2010general, choquet2015introduction}. Many physical predictions of GR have been experimentally validated over the last century, the most recent being the detection of gravitational waves \cite{abbott2016observation}. However, GR has some well known limitations, such as the breakdown of the equivalence principle at singularities, or the accelerating expansion of the Universe and the mystery of dark energy (responsible for this accelerated expansion). Therefore, many different modifications to GR have been proposed. Amongst these modified theories of gravity, scalar-tensor theories \cite{Fujii2003} are some of the most studied. There are two major reasons to study such theories; firstly, it is one of the simplest ways to modify GR, and secondly, some extensions of the Standard Model of particle physics predict the existence of scalar fields \cite{borodulin2022core, battaglieri2017arxiv}. This is further motivated by the experimentally confirmed existence of one scalar field in Nature, namely the Higgs field \cite{higgs1964broken, higgs1964bosons, atlas2012g}. Moreover, there are several proposed explanations for the nature of dark energy based on scalar-tensor theories \cite{Clifton2011, Joyce2014}. Some of these models predict a fifth force, which has not yet been detected on Earth or in the Solar System \cite{Dickey1994,Adelberger2003,Kapner2007}.

One way to mitigate this tension between theory and observation is by introducing a “screening mechanism” \cite{Burrage2017}, which allows the effects of the additional scalar fields to vary depending on the environment. Therefore, a screening mechanism would enable additional scalar fields to contribute to dark energy or dark matter while evading current experimental constraints on fifth forces. There are several models for such screened scalar fields with different types of screening mechanisms, such as chameleons \cite{Khoury2003,Khoury20032}; symmetrons \cite{Dehnen1992, Gessner1992, Damour1994, Pietroni2005, Olive2008, Brax2010,Hinterbichler2010,Hinterbichler2011}, whose fifth forces have been suggested as alternatives to particle dark matter \cite{Burrage2016_2,OHare:2018ayv,Burrage2018Sym,Kading2023}; galileons \cite{Dvali2000,Nicolis2008,Ali2012}; and environment-dependent dilatons \cite{Damour1994,Gasperini:2001pc,Damour:2002nv,Damour:2002mi,Brax:2010gi,Brax:2011ja,Brax2022}. Most of these models have been or are proposed to be tested in a zoo of different experiments and observations, for example, Refs.~\cite{Sakstein2016,Burrage2017,Pokotilovski:2012xuk,Pokotilovski:2013tma,Jenke2014,Copeland2014,Hamilton2015,Lemmel2015,Burrage:2015lya,Elder2016,Ivanov:2016rfs,Burrage2016,Jaffe2017,Brax:2017hna,Sabulsky:2018jma,Brax:2018iyo,Cronenberg2018,Jenke2020, Pitschmann:2020ejb,Qvarfort:2021zrl,Brax:2021wcv,Yin:2022geb,Betz:2022djh, Hartley:2019wzu,Fischer:2023koa,Fischer:2023eww,Fischer:2024coj,universe10030119}. Furthermore, in recent years, there have been initial attempts to study screened scalars as quantum fields \cite{Brax2018quantch,Kading1,Kading2,Kading:2023mdk}, and it was proposed to study screened scalar-tensor theories in analogue gravity simulations \cite{Hartley2019}.  

Additional proposals in the particular case of chameleon fields suggest that experiments which measure Casimir forces may also be used to constrain chameleon theories\footnote{Recently, a proposal has been made to use Casimir experiments to constrain symmetron models \cite{Elder:2019yyp}. However, in this article we will only consider chameleon fields and leave symmetrons to future investigation.} \cite{mota2007evading, brax2007detecting, brax2010tuning, brax2015casimir, almasi2015force,Brax:2022uiv, Haghmoradi:2024ytw}. From a theoretical point of view, a natural extension of the previous proposals then arises: if the chameleon field can be constrained by the static Casimir effect, then it might also be constrained by the dynamical Casimir effect. Thus, the aim of the present work is to study the DCE in the presence of a chameleon field, and to explore the relationship between the particle production and the chameleon field parameters. As a first step towards estimating the feasibility of constraining chameleon fields with the DCE, we consider a toy model with only the effect of the chameleon field and no gravity. Since the problem we want to solve in this work is that of a confined quantum field with moving boundaries, we will use the techniques developed in Ref.~\cite{barbado2021}.

The article is organised as follows. In Sec.~\ref{Screen}, we introduce screened scalar fields using the example of the chameleon mechanism; and, in Sec.~\ref{preliminaries}, we describe some relevant aspects of QFTCS applied to the DCE and, in particular, the method developed in Ref.~\cite{barbado2021}. Sec.\,\,\ref{DCE_cham} is the nuclear part of the article, where we obtain the main result, and analyse it both analytically and with a numerical example. We conclude in Sec.\,\,\ref{Conclusion}. In addition, in Appendix \ref{A_n}, we show the derivation of the normalisation constant of the cavity modes. We use natural units $\hbar = c =1$ throughout the article.
%%%%%%%%%%%%%%%%%%%%%%%%%%%%%%%%%%%%%%%%%%%%%%%%%%%%%%%%%%%%%%%%%%%%%

\section{Background}\label{Bckgd}

In this section, we give an overview of scalar-tensor theories of gravitation, in particular screened scalar fields and the chameleon model. In addition, we show schematically the usual approach to studying the DCE within the framework of QFTCS, and we then outline the techniques that are used in the present work. 
%%%%%%%%%%%%%%%%%%%%
\subsection{Screened scalar fields}\label{Screen}

The aim of scalar-tensor theories of gravitation is to study the modifications of GR due to an additional scalar field which is coupled to the metric tensor. A common way of performing such a coupling between a scalar field $\varphi$ and the metric tensor $g_{ \mu \nu }$ is through a conformal factor $A^2(\varphi)$, such that
\beq
\label{J_E_metrics}
\tilde{g}_{\mu\nu}=A^2(\varphi)g_{\mu\nu}.
\eeq
In this sense, scalar-tensor theories of gravity are defined up to a conformal transformation leading from one so-called conformal frame to another\footnote{A more general class of scalar-tensor field models can be constructed from the disformal transformation $g'_{\mu \nu} = A^2(\varphi)g_{\mu \nu } + B(\varphi) \partial_{\mu} \varphi \partial_{\nu} \varphi$ \cite{Bekenstein:1992pj}. For an overview of disformally coupled scalar field models, see Ref.~\cite{sakstein2014disformal}.}. These conformal frames are merely different mathematical formulations. Hence, the theoretical prediction for an observable quantity cannot be altered due to a change of conformal frame. The advantage is that some calculations might be easier to perform in one frame than in another. Two popular conformal frames are the Jordan frame - with a metric we denote $\tilde{g}_{\mu\nu}$ - and the Einstein frame denoted as $g_{\mu\nu}$.   

Even though the physical measurement cannot be changed, the physical interpretation can actually differ from one frame to another. For example, in the Jordan frame formulation, Einstein’s theory of gravity is modified in such a way that test particles follow different geodesics from those predicted in GR, while in the Einstein frame formulation, test particles still follow GR’s geodesics but are also subject to a gravity-like fifth force of Nature carried by the additional scalar field $\varphi$. The problem with such a prediction is that fifth forces are tightly constrained in our Solar System. An interesting way to solve this issue is given by so-called screening mechanisms. Such a mechanism allows the fifth force to be weak within our Solar System but cosmologically significant on intergalactic scales. As we describe in the Introduction, Sec.\,\,\ref{Intro}, there are several models for such screened scalar fields with different types of screening mechanisms such as the chameleon model, which will be presented in more detail in Sec.\,\,\ref{ch}.

\subsubsection{Einstein-frame action}
In this article, we consider a universe containing a free scalar test particle $\Phi$, which we denote as the  “matter”, with mass $m_\Phi$; and an additional scalar field $\varphi$ conformally coupling to the metric tensor. In the Einstein frame, this universe's action is schematically given by
\beq
\label{S}
S_{Universe} = S_{gravity} + S_m + S_{\varphi},
\eeq 
where $S_{gravity}$ is the usual Einstein-Hilbert gravitational action, $S_m$ the matter action, and $S_{\varphi}$ the action of the scalar field $\varphi$. Following Ref.~\cite{Kading1}, the conformal coupling to the metric tensor induces an interaction between $\Phi$ and $\varphi$, which in turn leads to a rescaling of the free field's mass by the conformal factor. Consequently, the Lagrangian matter density associated with the action $S_m$ is given by
\beq
\label{L_E}
\mathcal{L}_{m} = -\frac{1}{2}g^{\mu\nu}\partial_{\mu}\Phi\partial_{\nu}\Phi-\frac{1}{2}A^{2}\left(\varphi\right) m_\Phi^2\Phi^{2}.
\eeq
Subsequently, from the Euler-Lagrange equations, we obtain the equation of motion for the probe field $\Phi$:
\beq
\label{KG_cham}
g^{\mu\nu}\partial_{\mu}\partial_{\nu}\Phi-A^{2}\left(\varphi\right) m_\Phi^2 \Phi=0.
\eeq
Later, in Sec.~\ref{DCE_cham}, we will ignore gravity for simplicity and consequently set $g_{\mu\nu} = \eta_{\mu\nu}$.
%%%%%%%%
\subsubsection{Chameleon model} \label{ch}

A chameleon scalar field model has the defining property of coupling to matter in such a way that its effective mass increases with increasing local matter density. As its name suggests, the chameleon field adapts to its environment and becomes almost impossible to detect in regions of high matter density like our Solar System. The conformal coupling factor in Eq.~\eqref{J_E_metrics} of a chameleon is given by 
\beq
\label{ch_factor}
A^2(\varphi)= \rme ^{2\varphi/M},
\eeq
where $M$ is a mass scale which determines the strength of the chameleon-matter coupling. As is common practice when dealing with chameleons, we assume that $\varphi/M\ll1$. The Lagrangian density describing the chameleon field and associated to the action $S_{\varphi}$ in Eq.~(\ref{S}) is
\beq
\label{L_ch}
\mathcal{L}_{\varphi} = -\frac{1}{2} \left( \partial_{\varphi} \right)^2 -\frac{\Lambda ^{4+N}}{\varphi ^N} - \frac{\varphi}{M} \rho\,\,,
\eeq

\noindent where $N \in \mathbb{Z}^+ \cup  2 \mathbb{Z}^- \backslash \{-2\}$ distinguishes between different chameleon models; the parameter $\Lambda$ determines the strength of the self-interaction; and $\rho$ is the density of non-relativistic matter that the chameleon is interacting with. The sum of the last two terms in Eq.~\eqref{L_ch} results in an effective potential with a local minimum, and consequently a non-vanishing chameleon mass which increases with the matter density. Since the chameleon fifth force usually has a Yukawa-like suppression \cite{Khoury2003}, its range is the shorter  the larger the chameleon's mass. Consequently, in environments of sufficiently high density, the chameleon fifth force is effectively quite feeble, i.e., screened.  

Consider a static spherically symmetric source of radius $R$ and homogeneous density $\rho_{obj}$ immersed in a homogeneous medium of density $\rho_{bg}$. The field profile outside of this source, but still within an ambient Compton wavelength $(r<m_{bg}^{-1})$ with $m_{bg}$ being the chameleon's mass in the medium of density $\rho_{bg}$, is approximately given by \cite{Elder2016}
\beq
\label{motion_sol_C}
\varphi \simeq \varphi_{bg}-\frac{R}{r} \left( \varphi_{bg} - \varphi_{obj} \right),
\eeq
where $\varphi_{obj}$ is the value of the chameleon field inside the source and $\varphi_{bg}$ is the value of the chameleon field outside the source or the so-called background value. In the case of a large density contrast $\rho_{obj} \gg \rho_{bg}$, we can consider $\varphi_{bg} \gg \varphi_{obj}$. If the source is screened, then $\varphi_{obj}$ is actually the minimum of the chameleon within the source apart from a thin shell near the surface. Only the matter in this thin shell sources the chameleon fifth force in the exterior while the interior is not contributing. This is due to the short range of the fifth force in case of a large effective chameleon mass, and is known as the \emph{thin-shell effect}. In order to know if the chameleon field is screened or not, we define the shell thickness
\beq
\label{shell}
\Delta R = \frac{M \varphi_{bg}}{\rho_{obj}  R}.
\eeq
The object is said to be screened if $\Delta R \ll R$ or
\beq
\label{th_sh}
\frac{M \varphi_{bg}}{\rho_{obj} R^2} \ll 1.
\eeq
%

%%%%%%%%%%%%%%%%%%%
\subsection{Dynamical Casimir effect, quantum field theory and particle content}\label{preliminaries}

The usual approach to studying the DCE is to consider a free scalar field $\Phi$ in a one-dimensional\footnote{There are some authors that compute the DCE in three dimensions \cite{bialynicki2008dynamical, sassaroli1998dynamical, celeri2009action}. We consider the one-dimensional case because the transverse momentum does not appear, which simplifies the calculations. Moreover, experimental investigations of the DCE primarily work in one dimension \cite{wilson2011observation, lahteenmaki2013dynamical, svensson2018microwave}.} cavity with perfectly reflecting boundaries satisfying the Klein-Gordon equation
\beq
\label{KG_method}
g^{\mu \nu} \nabla_{\mu} \nabla_{\nu} \Phi - m_\Phi^2 \Phi - \xi \mathcal{R} \Phi = 0,
\eeq
where $m_\Phi \geq 0$ is the rest mass of the field, $g^{\mu \nu}$ is the spacetime metric, $\mathcal{R}$ its scalar curvature and $\xi \in \mathbb{R}$ is a coupling constant. Let us consider flat spacetime in inertial coordinates $\left(t,x\right)$. The boundaries of the cavity are moved during the time $t_{0} < t < t_{f}$. Since we are considering ideally reflecting boundaries, we impose Dirichlet vanishing boundary conditions
\beq
\label{Dir_conds_bg}
\Phi \left(t, x = x_{l} \left(t \right) \right) = \Phi \left(t, x = x_{r} \left(t \right) \right) = 0, 
\eeq
where the functions $x_{l}\left(t\right)$ and $x_{r}\left(t\right)$ determine the positions of the left and right boundaries for $t_{0} < t < t_{f}$, respectively. Before the boundaries move $(t < t_{0})$, we assume that the walls are static. For such initial conditions, the quantised field operator is decomposed as follows \cite{BD1984}: 
\beq 
\label{Q_field_bg}
\hat{\Phi} \left(t, x \right) = \underset{n}{\sum} \left[ \hat{a}_{n} \phi_{n} \left(t, x \right) + \hat{a}_{n}^{\dagger} \phi_{n}^{*} \left(t, x \right) \right],
\eeq 
where the mode functions $\phi_{n}\left(t,x\right)$ are solutions to the Klein-Gordon equation \eqref{KG_method}. In addition, $\hat{a}_{n}$ and $\hat{a}_{n}^{\dagger}$ are the bosonic annihilation and creation operators, respectively. Hence, the Fock space and vacuum state are defined in the canonical way. Two sets of mode solutions are related by a Bogoliubov transformation. In this way, the effects of the moving boundaries on the quantum field can be computed using a Bogoliubov transformation \cite{BD1984}, such that 
\begin{align}
    \tilde{\phi}_{m} &= \underset{n}{\sum} \left[ \alpha_{mn} \phi_{n} + \beta_{mn} \phi_{n}^{*} \right], \nonumber \\
    \tilde{a}_{m} &= \underset{n}{\sum} \left[ \alpha_{mn}^{*} \hat{a}_{n} + \beta_{mn}^{*} \hat{a}_{n}^{\dagger} \right], \label{a_transform}
\end{align}
where $\alpha_{mn}$ and $\beta_{mn}$ are called Bogoliubov coefficients. Note that if $\beta_{mn} \neq 0$, then the transformation of the annihilation operator of Eq.~\eqref{a_transform} contains creation operators. Therefore, the two vacua do not coincide. Hence, the $\beta$-coefficients quantify particle creation due to the transformation. Starting with a vacuum state, the average number of particles in mode $m$ after a Bogoliubov transformation is given by 
\beq 
\label{part_num}
\mathcal{N}_{m} = \underset{n}{\sum} \left| \beta_{nm} \right|^{2}. 
\eeq 

In general, the computation of the Bogoliubov coefficients is difficult. Thus, mathematical techniques and simplifications adapted to a specific problem make the computations manageable. For instance, the presence of symmetries like homogeneity or isotropy is convenient to obtain results on particle creation in cosmological models. In Refs.~\cite{barbado2020,barbado2021}, the authors developed a method to compute the Bogoliubov transformation experienced by a confined quantum scalar field in a globally hyperbolic spacetime due to the changes in the geometry and/or the confining boundaries. The second part \cite{barbado2021} extends the method to cases in which the timelike boundaries of the spacetime do not remain static in any synchronous frame. This method is especially useful in the presence of resonances of the field modes due to small perturbations of the metric and/or the motion of the cavity boundaries. This is because in these cases, the Bogoliubov coefficients take the following simple expressions
\begin{align}
\alpha_{n n} (t_{\mathrm{f}}, t_0) \approx &\ 1; \nonumber \\
\alpha_{n m} (t_{\mathrm{f}}, t_0) \approx &\ \varepsilon \int_{t_0}^{t_{\mathrm{f}}} \rmd t \,\rme^{ -\rmi (\omega^0_n - \omega^0_m) t} \Delta \hat{\alpha}_{n m} (t), \quad n \neq m; \label{perturbation_alpha} \\
\beta_{n m} (t_{\mathrm{f}}, t_0) \approx &\ \varepsilon \int_{t_0}^{t_{\mathrm{f}}} \rmd t \,\rme^{ - \rmi (\omega^0_n + \omega^0_m) t} \Delta \hat{\beta}_{n m} (t); \label{perturbation_beta}
\end{align}
where $\varepsilon \ll 1$ is a small parameter that characterises the perturbation of the confined field (e.g. oscillation amplitude), and $\omega_{n}^{0}$ are the mode frequencies for the static problem $( \varepsilon = 0)$. For Dirichlet boundary conditions,
\begin{align}
\Delta &\hat{\alpha}_{n m} (t) \equiv \ \rmi \int_{\Sigma^0} \rmd V^0\ [\prescript{-}{m}{\hat{\Delta}}(t) \Psi^0_n] \Psi^0_m  - \rmi \int_{\partial \Sigma_{0}} \rmd S^0\ \Delta x(t) \left(\mathbf{n} \cdot \nabla_{h^0} \Psi_n^0 \right)\left(\mathbf{n} \cdot \nabla_{h^0} \Psi_m^0\right), \label{ssd_alpha_hat_moving} \\
\Delta &\hat{\beta}_{n m} (t) \equiv \ -\rmi \int_{\Sigma^0} \rmd V^0\ [\prescript{+}{m}{\hat{\Delta}}(t) \Psi^0_n] \Psi^0_m  + \rmi \int_{\partial \Sigma_{0}} \rmd S^0\ \Delta x(t) \left(\mathbf{n} \cdot \nabla_{h^0} \Psi_n^0 \right)\left(\mathbf{n} \cdot \nabla_{h^0} \Psi_m^0 \right). \label{ssd_beta_hat_moving}
\end{align}
Here, $\Sigma^{0}$ is a fixed spatial hypersurface, around which the perturbation occurs, with volume element $dV^{0}$; boundary $\partial \Sigma^{0}$; boundary surface element $dS^{0}$; proper distance $\varepsilon \Delta x(t)$ between the boundary $\partial \Sigma_{t}$ and the fixed boundary $\partial \Sigma^{0}$; and connection $\nabla_{h^0}$ associated to the static metric~$h^0_{ij}$. $\,_{m}^{+} \hat{\Delta}(t)$ are linear operators determined by their actions on the mode basis $\{\Psi^0_n\}$ as defined in Eq.~(50) of Ref.~\cite{barbado2021}. 

The Bogoliubov transformation differs maximally from the identity just by terms of first order in~$\varepsilon$, except for the cases where there are resonances. If the perturbation considered contains some characteristic frequency~$\omega_{\mathrm{p}}$, such that it coincides with some difference between the frequencies of two modes, $\omega_{\mathrm{p}} = \omega^0_n - \omega^0_m$, then the corresponding coefficient~$\alpha_{n m} (t_{\mathrm{f}}, t_0)$ grows linearly with the time difference~$t_{\mathrm{f}} - t_0$ and eventually grows to be a non-perturbative correction. Respectively, if the characteristic frequency coincides with some sum between the frequencies of two modes, $\omega_{\mathrm{p}} = \omega^0_n + \omega^0_m$, then the coefficient~$\beta_{n m} (t_{\mathrm{f}}, t_0)$ grows linearly in time. The duration of the perturbation $\Delta t$ should be such that $1 \ll \omega_{\mathrm{p}} \Delta t \ll 1/\varepsilon$. This is because the period of time should be reasonably larger than the inverse of the frequency being described, but on the other hand, one should keep higher order terms in~$\varepsilon$ significantly smaller than the first order term to ensure the validity of the perturbative computation.

%%%%%%%%%%%%%%%%%%%%%%%%%%%%%%%%%%%%%%%%%%%%%%%%%%%%%%%%%%%%%%%%%%%%%%%%%%

\section{Dynamical Casimir effect in a spacetime with a screened scalar field}\label{DCE_cham}

In this section, we study the toy model of a DCE for a minimally coupled massive quantum scalar field in a spacetime affected by a chameleon field $\varphi=\varphi \left( \mathbf{x} \right)$. Let us consider the spacetime metric to be the Minkowski metric $\eta_{\mu\nu}$ and a quantum field trapped inside an effectively one-dimensional cavity\footnote{The reduction from a three spatial dimensional problem is done by assuming that two cavity dimensions are much smaller than the third one, such that the system can be effectively treated in one spatial dimension. This is due to the direction of the chameleon field gradient being radial, and therefore no significant effects occurring in the transverse directions.} of average proper length $L$. The cavity is placed at a distance $d$ to a sphere of radius $R$, which acts as a source for the chameleon force. We consider coordinates centered on the sphere, where $x$ is the radial distance to the center of the sphere. The boundaries of the cavity are placed at $x_{l}$ (left) and $x_{r}$ (right), and the right boundary oscillates with frequency $\Omega$ and amplitude $\varepsilon L\ll L$, such that the cavity is oscillating as: 
\beq
\label{boundaries}
x_{l}=s, \quad x_{r} \left(t\right) = s +L \left[1+\varepsilon\sin\left(\Omega t\right)\right],
\eeq
where $s=R+d$. We impose Dirichlet boundary conditions on the scalar field at the boundaries and ignore the gravitational field of the chameleon source mass. This toy model will help us to understand the qualitative behaviour of a confined quantum field with moving boundaries in a spacetime with a screened scalar field. 

Since the Minkowski metric $\eta_{\mu \nu}$ is a synchronous frame, we can use the method displayed in Ref.~\cite{barbado2021} right away. Focusing on the small perturbations regime for the problem under consideration, the quantities needed to compute Eqs.~\eqref{ssd_alpha_hat_moving} and \eqref{ssd_beta_hat_moving} are
\begin{align}
 \,_{m}^{\pm}\hat{\Delta} &= 0; \label{sup_op_computed} \\
 \Delta x (s) = 0,\quad \Delta x (s+L) &= L\sin(\Omega t). \label{Delta_bounds}
\end{align}
Using Eq.~\eqref{KG_cham}, the static spatial eigenvalue equation (34) of Ref.~\cite{barbado2021} and the boundary conditions read  
\begin{align}
\label{eigeq_Chris}
    \partial_{x}^{2} \Psi_{n}^{0} + \left[ \left( \omega_{n}^{0} \right)^{2} - A^{2} \left( \varphi \right) m_\Phi^{2} \right] \Psi_{n}^{0} &= 0, \nonumber \\
    \Psi_{n}^{0} \left(x_l = s\right) = \Psi_{n}^{0} \left(x_r (0) = s + L\right) &= 0 .
\end{align}
%
%%%%%%%%%%%%%%%%%%%%%%%%%%%%%%

\subsection{Solution to the static eigenvalue equation}\label{sol_eigeq_sec}

Since we have assumed $\varphi /M \ll 1$, see Sec.~\ref{ch}, the chameleon coupling function in Eq.~\eqref{ch_factor} can be approximated by  
\beq
\label{A_approx}
A^{2} \left( \varphi \right) = 1 + 2 \frac{ \varphi \left( x \right)}{M} + \mathcal{O} \left( \frac{ \varphi^2}{M^2} \right). 
\eeq
The chameleon profile given in Eq.~\eqref{motion_sol_C} is a function of $1/r$ with $r>0$. Thus, assuming that the cavity is sufficiently far from the source mass center, i.e. $L \ll s$, it is possible to linearise the field profile within the cavity $x\in\left[x_l,x_r\right]$, such that 
\beq
\label{motion_C_lin}
\varphi \left( x \right) \approx \varphi_{bg} - 2 \varphi_{bg} \frac{R}{s} + \varphi_{bg} \frac{R}{s^{2}} x.
\eeq
Substituting Eq.~\eqref{motion_C_lin} in the eigenvalue equation \eqref{eigeq_Chris}, we have 
\beq
\label{eigeneq_lin}
\partial_{x}^{2} \Psi_{n}^{0} - 2 \frac{\varphi_{bg}}{M} \frac{R}{s^{2}} x m_\Phi^{2} \Psi_{n}^{0} + \left[ \left( \omega_{n}^{0} \right)^{2} - \left[ 1 + 2 \frac{\varphi_{bg}}{M} - 4 \frac{ \varphi_{bg}}{M} \frac{R}{s} \right] m_\Phi^{2} \right] \Psi_{n}^{0} = 0.
\eeq
Following the technique presented in Ref.~\cite{sorge2005casimir}, we define the quantities  
\begin{align}
a &:= 2 \frac{ \varphi_{bg}}{M} \frac{R}{s^{2}} m_\Phi^{2}, \label{a}  \\ 
b &:= 1 + 2 \frac{ \varphi_{bg}}{M} - 4\frac{ \varphi_{bg}}{M} \frac{R}{s}, \label{b} \\
\lambda_{n}^{2} &:= \left( \omega_{n}^{0} \right)^{2} - b m_\Phi^{2}. \label{lambda}
\end{align}
Furthermore, we introduce a new variable 
\beq
\label{u}
u := u_n \left( x \right) = \left( \frac{ \lambda_{n}^{2}}{a} - x \right) \left( a \right)^{1/3}.
\eeq
From here on, we omit the explicit dependence of $u$ on $n$ for simplicity in the notation. Then the eigenvalue equation \eqref{eigeneq_lin} can be rewritten as
\beq
\label{eigeq_u}
\partial_{u}^{2}\Psi_{n}^{0}+u\Psi_{n}^{0}=0,
\eeq
which is an Airy differential equation. The solution of Eq.~\eqref{eigeq_u} is given by means of Bessel functions:
\beq
\label{Bessel}
\Psi_{n}^{0} \left( u \right) = \sqrt{u} \left[ c_{1} J_{1/3} \left( \frac{2}{3} u^{3/2} \right) + c_{2} J_{-1/3} \left( \frac{2}{3} u^{3/2} \right) \right].
\eeq
To progress further in this derivation, we must assume that $u \gg 1$. Therefore, we can apply the asymptotic form of Eq.~\eqref{Bessel} as seen in Ref.~\cite{sorge2005casimir}, such that  
\beq
\label{Psi_asymptotic}
\Psi_{n}^{0}\left(u\right)=\mathcal{A}_{n}u^{-1/4}\sin\left(\frac{2}{3}u^{3/2}+\phi\right),
\eeq
with $\mathcal{A}_{n}$ and $\phi$ being constants. While $u \gg 1$ is not globally true, we show in Sec.~\ref{analysis_sec} that the region of the parameter space of the considered chameleon models, for which this assumption can be applied, is largely unconstrained by experiments.

To guarantee that the field satisfies the Dirichlet boundary conditions, we require that
\beq
\label{n_pi}
\frac{2}{3}\left[u^{3/2}\left(x_{r}\right)-u^{3/2}\left(x_{l}\right)\right]=n\pi,\qquad n\in\mathbb{N}.
\eeq
The next approximation we make is that the variation of $u$ within the cavity is small. This lets us linearise $\left[u\left(x\right)\right]^{3/2}$ at the point $s$, 
\beq
\label{expansion_u}
\left[ u \left( x \right) \right]^{3/2} \approx \left[ u \left( s \right) \right]^{1/2} \left[ u \left( s \right) + \frac{3}{2} \left[ u' \left( s \right) \right] \left( x - s \right) \right].
\eeq
Then Eq.~\eqref{n_pi} becomes
\beq
\label{n_pi_expand}
\left[ u \left( s \right) \right]^{1/2} u' \left( s \right) L = n\pi .
\eeq
Substituting Eqs.~\eqref{u} and \eqref{lambda} in Eq.~\eqref{n_pi_expand}, we obtain 
\beq
\label{omegas}
\left( \omega_{n}^{0} \right)^{2} = k_{n}^{2} + m_\Phi^{2} \left( 1 + 2 \frac{\varphi_{bg}}{M} \left[ 1 - \frac{R}{s} \right] \right),
\eeq
where $k_{n}=\frac{n\pi}{L}$. Note that if the chameleon field is turned off, we recover the usual frequencies of the static problem in flat spacetime for a massive field \cite{BD1984}
\beq
\label{omega_static}
\omega_{n_{\text{f}}}^{0}=\sqrt{k_n^2 + m_\Phi^2}.
\eeq
From Eq.~\eqref{Psi_asymptotic} and the boundary condition $\Psi_{n}^{0}\left(x_{l}\right)=0$, we see that
\beq
\label{phi_phase}
\phi=-\frac{2}{3}u^{3/2}\left(x_{l}\right).
\eeq
Applying the normalisation condition given in Eq.~(36) of Ref.~\cite{barbado2021}, we obtain 
\beq
\label{A_cte} 
\mathcal{A}_{n} = \left[ 2 \omega_{n}^{0} \int_{x_{l}}^{x_{r}} \rmd x \,u^{-1/2} \left( x \right) \sin^{2} \left( \frac{2}{3} u^{3/2} \left( x \right) - \frac{2}{3} u^{3/2} \left( x_{l} \right) \right) \right]^{-1/2}.
\eeq
Using Eq.~\eqref{expansion_u} and doing another linearisation of $u^{-1/2}\left(x\right)$ at the point $s$, we obtain after some algebra
\beq
\label{A_cte_L} 
\mathcal{A}_{n} = \frac{2 \left( k_{n}^{2} \right)^{3/4}}{ \left( \omega_{n}^{0} L \right)^{1/2} \left( a \right)^{1/6} \left( 4 k_{n}^{2} + a L \right)^{1/2}}.
\eeq
The full derivation of this normalisation constant can be found in the Appendix \ref{A_n}.
%%%%%%%%%%%%%%%%%%%%%%%%%%%%%%%%%%%%

\subsection{Bogoliubov coefficients}\label{bogo_sec}

In order to compute the Bogoliubov coefficients in Eqs.~\eqref{perturbation_alpha} and \eqref{perturbation_beta}, we first need to compute the quantities in Eqs.~\eqref{ssd_alpha_hat_moving} and \eqref{ssd_beta_hat_moving}. Note that the first integrals of Eqs.~\eqref{ssd_alpha_hat_moving} and \eqref{ssd_beta_hat_moving} vanish since the operator $\,_{m}^{\pm}\hat{\Delta}$ is zero (as seen in Eq.~\eqref{sup_op_computed}). For the second integrals, we substitute Eq.~\eqref{Delta_bounds}. Since we are considering one spatial dimension, the “surface integral” is simply the evaluation of the integrand at the two static boundaries, such that 
\begin{align}
\label{alphabeta_hat_Dir}
\Delta \hat {\alpha}_{nl} &= - \rmi L \sin \left( \Omega t \right) \left( -1 \right)^{n+l} \mathcal{A}_{n} \mathcal{A}_{l} \left( a \right)^{1/3} 
\left[ \left( \lambda_{n}^{2} - a x_{r} \right) \left( \lambda_{l}^{2} - a x_{r} \right) \right]^{1/4} \nonumber \\
&= - \Delta \hat{\beta}_{nl}. 
\end{align}
Substituting Eq.~\eqref{A_cte_L} in Eq.~\eqref{alphabeta_hat_Dir} and considering that the oscillation frequency of the boundary coincides with the difference of the mode frequencies, that is $\Omega = \left| \omega_{n}^{0} - \omega_{l}^{0} \right|$, then we see that the $\alpha$-coefficients from Eq.~\eqref{perturbation_alpha} are given by
\beq
\label{alpha_Dir_r}
\alpha_{nl} \left(t_{f},t_{0}\right) \approx - \varepsilon \frac{2\left( -1 \right)^{n+l} \left( t_{f}-t_{0} \right)}{ \left( \omega_{n}^{0} \omega_{l}^{0} \right)^{1/2}} \frac{ \left[ k_{n}^{2} k_{l}^{2} \right]^{3/4} \left[ k_{n}^{2} k_{l}^{2} - aL \left( k_{n}^{2} + k_{l}^{2} \right) + a^{2} L^{2} \right]^{1/4}}{ \left[ 16 k_{n}^{2} k_{l}^{2} + aL \left( 4k_{n}^{2} + 4k_{l}^{2} \right) + a^{2} L^{2} \right]^{1/2}}.
\eeq
If, instead, the oscillation frequency of the boundary coincides with the sum of the mode frequencies, that is $\Omega = \omega_{n}^{0} + \omega_{l}^{0}$, then Eq.~\eqref{perturbation_beta} is
\beq
\label{beta_Dir_r}
\beta_{nl} \left(t_{f},t_{0}\right) \approx \varepsilon \frac{2 \left( -1 \right)^{n+l} \left(t_{f}-t_{0}\right)} {\left( \omega_{n}^{0} \omega_{l}^{0} \right)^{1/2}} \frac{ \left[ k_{n}^{2} k_{l}^{2} \right]^{3/4} \left[ k_{n}^{2} k_{l}^{2} - aL \left( k_{n}^{2} + k_{l}^{2} \right) + a^{2} L^{2} \right]^{1/4}}{ \left[ 16 k_{n}^{2} k_{l}^{2} + aL \left( 4k_{n}^{2} + 4k_{l}^{2} \right) + a^{2} L^{2} \right]^{1/2}}.
\eeq
Eqs.~\eqref{alpha_Dir_r} and \eqref{beta_Dir_r} are the general results of this work. Recall that these coefficients are obtained in the presence of resonances where the corresponding coefficient grows linearly with time. Hence, after enough time the effect becomes significant and non-perturbative. 

%%%%%%%%%%%%%%%%%%%%%%%%%%%%%%

\subsection{Analysis}\label{analysis_sec}

If we expand Eq.~\eqref{beta_Dir_r} around the small parameter $\frac{\varphi_{bg}}{M}$ up to first order (since the second order is negligible, see Eq.~(\ref{A_approx})), we obtain
\beq 
\label{beta_expanded}
\beta_{nl} \left(t_{f},t_{0} \right) \approx \frac{\left( -1 \right)^{n+l} \varepsilon k_{n} k_{l} \left( t_{f} - t_{0} \right)} { 2 \left( \omega_{n_{\text{f}}}^{0} \omega_{l_{\text{f}}}^{0} \right)^{1/2}} \left\{ 1 + B_{nl} + \mathcal{O} \left( \frac{\varphi_{bg}^{2}}{M^{2}}\right) \right\}.
\eeq
We see that the zeroth order approximation, when the chameleon field is turned off, gives the usual coefficients of the DCE in Minkowski spacetime \cite{ji1997production, sabin2014, barbado2021}. Thus, the zeroth order approximation exhibits the familiar resonance behaviour in the $\beta$-coefficients. The first order approximation is given by 
\beq
\label{delta}
B_{nl} = \frac{m_\Phi^{2}}{2} \frac{\varphi_{bg}}{M} \left[ -\left( \frac{1}{\left[ \omega_{n_{\text{f}}}^0 \right]^{2}} + \frac{1}{\left[ \omega_{l_{\text{f}}}^0 \right]^{2}} \right) + \frac{R}{s} \left( \frac{1}{ \left[ \omega_{n_{\text{f}}}^0 \right]^{2}} + \frac{1}{\left[ \omega_{l_{\text{f}}}^0 \right]^{2}} \right) - \frac{3}{2} \frac{RL}{s^{2}} \left( \frac{k_{n}^{2} + k_{l}^{2}}{k_{n}^{2} k_{l}^{2}} \right) \right].
\eeq
Eq.~\eqref{delta} gives a novel contribution due to the chameleon field, where the first term is a constant independent of the geometry. The second contribution depends on the position of the cavity in relation to the chameleon source, while the third term depends on the length of the cavity and tells us about the strength of the chameleon gradient between the two ends of the cavity, reminiscent of the structure of a linearised Newtonian gravitational potential. Since we are already in the resonance regime $\Omega = \omega_{n}^{0} + \omega_{l}^{0}$, the sum in the number of particles in Eq.~\eqref{part_num} disappears, such that the average particle number is given by  
\beq
\label{particleN}
\left| \beta_{nl} \left( t_{f},t_{0} \right) \right|^{2} \approx \frac{ \varepsilon^{2} k_{n}^{2} k_{l}^{2} \left(t_{f}-t_{0}\right)^{2}} { 4 \left( \omega_{n_{\text{f}}}^0 \omega_{l_{\text{f}}}^0 \right)} \left\{ 1 + 2B_{nl} \right \}.
\eeq
To see how the chameleon contribution affects the $\beta$-coefficients and thus the particle number, let us consider that the cavity and the chameleon source are inside a vacuum chamber of radius $R_{vac}$. We plot the contours for $B_{nl}$ using the parameters shown in Tab.~\ref{tab_numbers}. Note that we consider a kaon $K^0$ as the massive quantum scalar field in our toy model. 
\begin{table}[]
    \centering
    \begin{tabular}{ c c c }
    \hline 
       \textbf{Parameter}  & \textbf{Symbol} & \textbf{Value} \\
     \hline \hline  
       Cavity length  & $L$ & $50 ~\mathrm{eV}^{-1}$ ($ \sim 10^{-5}$ m) \\
       Cavity field mass & $m_\Phi$ & $497 \times 10^6$ eV (mass of kaon $K^0$) \\
       Source mass radius  & $R$ & $1000 ~\mathrm{eV}^{-1}$ ($2 \times 10^{-4}$ m) \\
       Distance from center of source mass to cavity  & $s$ & $5000 ~\mathrm{eV}^{-1}$ ($ \sim 10^{-3}$ m) \\
       Vacuum chamber radius  & $R_{vac}$ & $2.5 \times 10^5 ~\mathrm{eV}^{-1}$ ($5 \times 10^{-2}$ m) \\
    \hline    
    \end{tabular}
    \caption{Parameters used to compute the chameleon contribution to the particle content of the confined quantum field.}
    \label{tab_numbers}
\end{table}
In order to obtain the chameleon background value, we use the relation given in Refs.~\cite{Hamilton2015, Elder2016}
\beq
\label{ch_bg}
\varphi_{bg} \left( \Lambda \right) = \xi \left( N \left( N + 1 \right) \Lambda^{4+N} R_{vac}^{2} \right)^{1/N+2},
\eeq 
where $\xi$ is a "fudge factor" largely insensitive to $N$, $\Lambda$ and $M$, as well as to the assumed chamber geometry. Here, we assume the conservative value of $\xi = 0.55$ \cite{Elder2016}. We consider the chameleon models $N = 1$ and $N = -4$, which are the most studied ones \cite{Burrage2017}. For the chameleon model with $N = -4$, the Lagrangian in Eq.~\eqref{L_ch} changes to
\beq
\label{L_ch4}
\mathcal{L} = - \frac{1}{2} \left( \partial \varphi \right)^2 - \frac{\lambda}{4!} \varphi^4 - \frac{\varphi}{\mathcal{M}} \rho.
\eeq
Hence, in this case, Eq.~\eqref{ch_bg} is given by
\beq
\label{ch_bg4}
\varphi_{bg} = \xi \sqrt{\frac{2}{\lambda}} \frac{1}{R},
\eeq
where $\lambda = \left(\Lambda / \Lambda_{DE} \right)^4$ and $\Lambda_{DE}=2.4$ meV is the dark energy scale \cite{Burrage2017}.
\begin{figure}
    \centering
    \begin{minipage}{0.48\textwidth}
    \subfigure[]{\includegraphics[width=0.98\textwidth]{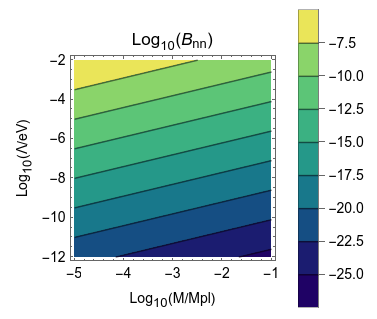}}
    \label{fig:B_nn3}
    \end{minipage}\hfill
    \begin{minipage}{0.48\textwidth}
    \subfigure[]{\includegraphics[width=0.98\textwidth]{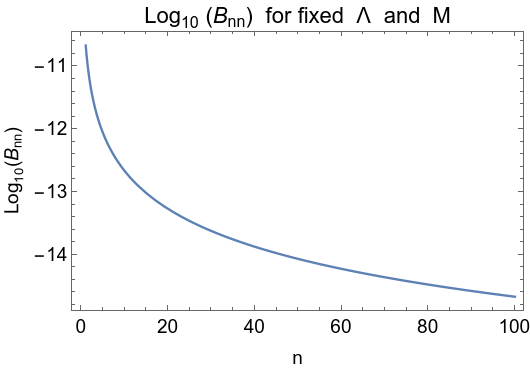}}
    \label{fig:B_nn2}
    \end{minipage}\hfill
\caption{Plots of the logarithmic values of the chameleon contribution to the $\beta$-coefficients and the particle number for the chameleon model $N = 1$. Figure (a) shows the contour plot of such a contribution as a function of $\Lambda$ and $M$ for fixed the quantum number $n = 1$. For this part of the parameter space, $u \gg 1$ is true. Plot (b) shows the contribution as a function of the quantum number $n$ for $\Lambda = 1 \times 10^{-3}$ eV and $M = 1 \times 10^{-1} M_{Pl}$.}
\label{fig:cham1}
\end{figure}

Fig.~\ref{fig:cham1} shows the chameleon contribution to the Bogoliubov coefficients and consequently to the particle number for the chameleon model $N = 1$, where, more precisely, Fig.~\ref{fig:cham1}a depicts the chameleon contribution as a function of $\Lambda$ and $M$. The parameter $M$ is essentially unconstrained but probably below the reduced Planck mass $M_{Pl} \approx 2.4 \times 10^{18}$ GeV \cite{Hamilton2015}. Here, we use a subset of the parameter spaces shown in Refs.~\cite{Hartley:2019wzu, Qvarfort:2021zrl}, where the assumption made in Eq.~\eqref{u} is fulfilled, namely $u \gg 1$ \footnote{With the parameters of Tab.~\ref{tab_numbers}, all the approximations made in this work are fulfilled, namely $\varphi/M \ll 1$ and $\varepsilon L \ll L \ll s \ll R_{vac} \ll m_{bg}^{-1}$.}. Fig.~\ref{fig:cham1}b shows the chameleon contribution as a function of the cavity mode number $n$. Note that the chameleon contribution is stronger for the upper left corner in Fig.~\ref{fig:cham1}a. In addition, also note that, for fixed $\Lambda$ and $M$, the chameleon contribution is the strongest for the quantum number $n=1$, but decays with increasing $n$.
\begin{figure}
    \centering
    \begin{minipage}{0.48\textwidth}
    \subfigure[]{\includegraphics[width=0.98\textwidth]{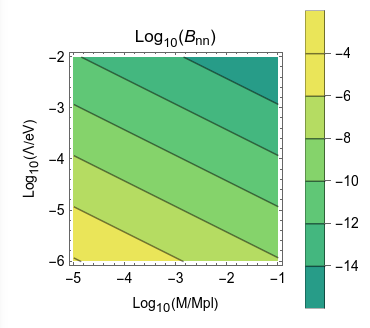}}
    \label{fig:4B_nn3}
    \end{minipage}\hfill
    \begin{minipage}{0.48\textwidth}
    \subfigure[]{\includegraphics[width=0.98\textwidth]{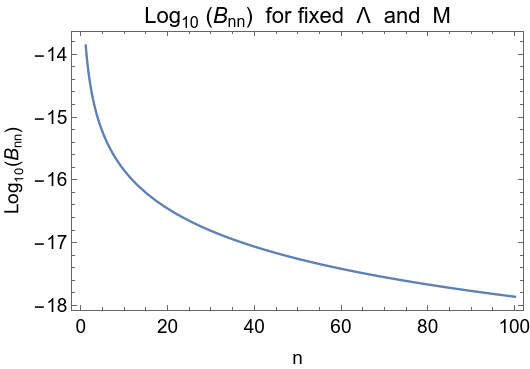}}
    \label{fig:4B_nn2}
    \end{minipage}\hfill    
\caption{Plots of the logarithmic values of the chameleon contribution to the $\beta$-coefficients and the particle number for the chameleon model $N = -4$. Plot (a) shows the contribution as a function of $\Lambda$ and $M$ for the fixed quantum number $n = 1$. For this part of the parameter space, $u \gg 1$ is true. Plot (b) shows the contribution as a function of the quantum number $n$ for $\Lambda = 1 \times 10^{-3}$ eV and $M = 1 \times 10^{-1} M_{Pl}$.}
\label{fig:cham4}
\end{figure}

Furthermore, in Fig.~\ref{fig:cham4}, we present the chameleon contribution to the Bogoliubov coefficients and consequently to  the particle number for the chameleon model $N = -4$. Fig.~\ref{fig:cham4}a shows the chameleon contribution as a function of $\Lambda$ and $M$, and Fig.~\ref{fig:cham4}b depicts it as a function of the cavity mode number $n$. Note that, in contrast to Fig.~\ref{fig:cham1}a, the chameleon contribution is stronger in the lower left corner in Fig.~\ref{fig:cham4}a, while, for fixed $\Lambda$ and $M$, it behaves in the same way as it did for the chameleon model $N = 1$.

This difference in behaviour can be understood by examining the chameleon-induced force. The acceleration experienced by a test particle due to a chameleon field near a spherical source mass is given by Ref.~\cite{Khoury2013}
\beq
\mathbf{a} = - \nabla \ln A\left(\varphi \right) = - \frac{\nabla{\varphi}}{M} = -\frac{R}{M r^{2}} \left(\varphi_{bg} - \varphi_{obj}\right) \mathbf{r},
\eeq
where $A$ is the conformal coupling factor defined in Eqs.~(\ref{J_E_metrics}) and (\ref{ch_factor}), and we have used the chameleon field profile in Eq.~(\ref{motion_sol_C}). It can immediately be seen that a smaller $M$ results in a stronger force. When comparing Eqs.~(\ref{ch_bg}) and (\ref{ch_bg4}), we see that the chameleon field scales oppositely with $\Lambda$ for $N=1$ and $N=-4$ models, i.e., $\varphi_{bg}$ increases with increasing $\Lambda$ for $N=1$ but decreases for $N=-4$.

Therefore, we come to the natural conclusion that the chameleon field effect on the cavity particle production is the strongest where the chameleon-induced force is also the strongest. In both considered cases for $N$, we find a reduction in the particle number due to the presence of the chameleon scalar field.   

%%%%%%%%%%%%%%%%%%%%%%%%%%%%%%%%%%%%%%%%%%%%%%%%%%%%%%%%%%%%%%%%%

\section{Conclusions}
\label{Conclusion}

In this paper, we have shown the effect of a chameleon field on the number of particles created in a massive quantum scalar field by the DCE. We have considered an effectively one-dimensional cavity, with one of its boundaries allowed to move, placed near a chameleon source mass. Since the chameleon field is coupled to the mass of the quantum field, the Klein-Gordon equation and, in particular, the Lagrange-Beltrami operator acting on the quantum field in a spatial hypersurface, are affected by the chameleon field. We then computed the Bogoliubov coefficients in the presence of parametric resonances using the techniques developed in Ref.~\cite{barbado2021}. For this computation, we have linearised the chameleon field profile, in analogy to studies of linearisations of the Newtonian potential \cite{celeri2009action}. As expected, when the chameleon is turned off, the Bogoliubov coefficients are those of the DCE in Minkowski spacetime. Finally, we have analysed how the particle content is affected by the presence of the chameleon field. We showed that the mean number of created particles is diminished by the presence of the chameleon field, and we gave representative numerical estimates for how the particle content is affected depending on the choice of the chameleon model, the parameters of the model, and the mode number of the quantum field.

This work can also be seen as an extension of the method presented in Refs.~\cite{barbado2020,barbado2021} since, for the first time, we were effectively considering a spatially dependent mass, which we can define as $\tilde{m}_\Phi \left( x \right) := A^{2} \left( \varphi \left( x \right) \right) m_\Phi$.  

To our knowledge, this article is the first work on the effect of screened scalar fields on particle creation. In the future, it will be interesting to create a more realistic study, also taking into account the gravitational field and not linearising the chameleon field. Besides, other screened scalar field models could also be studied in the same way. In addition, we leave for future work the study of entanglement between modes, their relation to the chameleon parameters, and the implementation of quantum metrology to estimate and constrain chameleon parameters. 

%%%%%%%%%%%%%%%%%%%%%%%%%%%%%%%%%%%%%%%%%%%%%%%%%%%%%%%%%%%%%%%%%

\begin{acknowledgments}
The authors are grateful to Luis C. Barbado, Tupac Bravo, Jes\'{u}s DelOlmo-M\'{a}rquez, Dennis R\"atzel and Jan Kohlrus for useful discussions. A.~L.~B.\ recognises support from CONAHCyT ref:579920/410674. C.~K. is supported by the Austrian Science Fund (FWF): P 34240-N. This publication is based upon work from COST Action COSMIC WISPers CA21106, supported by COST (European Cooperation in Science and Technology).
\end{acknowledgments}

\appendix

%%%%%%%%%%%%%%%%%%%%%%%%%%%%%%%%%%%%%%%%%%%%%%%%%%%%%%%%%%%%%%%%%

\section{Derivation of the normalisation constant} \label{A_n}

In order to compute the normalisation constant in Eq.~\eqref{A_cte}, we use Eq.~\eqref{expansion_u} and another linearisation of $u^{-1/2} \left( x \right)$ at the point $s$. Thus, the integral of the constant in Eq.~\eqref{A_cte} is 
\begin{align}
\label{I}
    I & = \int_{x_{l}}^{x_{r}} \rmd x \, u^{-1/2} \left( x \right) \sin^{2} \left( \frac{2}{3} u^{3/2} \left( x \right) - \frac{2}{3} u^{3/2} \left(x_{l} \right) \right) \nonumber \\
    & = \int_{x_{l}}^{x_{r}} \rmd x \left( \left[ u \left( s \right) \right]^{-1/2} \left( 1 - \frac{1}{2} \left[ u \left( s \right) \right]^{-1} u' \left( s \right) \left( x - s \right) \right) \right)  \nonumber \\
    & \hphantom{=+} \cdot \sin^{2} \left( \frac{2}{3} \left[ u \left( s \right) \right]^{1/2} \left[ \frac{3}{2} u'\left( s \right) x - \frac{3}{2} u'\left( s \right) x_{l} \right] \right).
\end{align}
Hence, 
\begin{align}
\label{I2}
    I & = \left[ u \left( s \right) \right]^{-1/2} \left\{ \frac{ \left( x-s \right)}{2} \left( 1 + \frac{1}{2} s u^{-1} \left( s \right) u' \left( s \right) \right) - \frac{1}{2} u^{-1} \left( s \right) u' \left( s \right) \left( \frac{ \left( x^{2} - s^{2} \right)}{4} \right) \right\} _{x_{l}}^{x_{r}} \nonumber \\
    & = \left[ u \left( s \right) \right]^{-1/2} \frac{L}{2} \left\{ 1 + \left( \frac{s}{2} - \frac{2s+L}{4} \right) u^{-1} \left( s \right) u' \left( s \right) \right\}.
\end{align}
Substituting Eq.~\eqref{u} in Eq.~\eqref{I2}, we obtain 
\begin{align}
\label{I3}
    I &= \left[ \left( \frac{ \lambda_{n}^{2}}{a} - s \right) \left( a \right)^{1/3} \right]^{-1/2} \frac{L}{2} \left\{ 1 + \left( \frac{s}{2} - \frac{2s+L}{4} \right) \left[ u \left( s \right) \right]^{-1} u'\left( s \right) \right\} \nonumber \\
    &= \frac{\left( a \right)^{1/2}}{\left[ \lambda_{n}^{2} - a s \right]^{1/2} \left( a \right)^{1/6}} \frac{L}{2} \left\{ 1 + \frac{L}{4} \left( \frac{a}{\left( \lambda_{n}^{2} - a s \right)} \right) \right \}.
\end{align} 
Replacing Eq.~\eqref{I3} in Eq.~\eqref{A_cte}, we have that
\beq
\label{An1}
\mathcal{A}_{n} = \left[ 2 \omega_{n}^{0} \frac{ \left( a \right)^{1/3}}{ \left[ \lambda_{n}^{2} - a s \right]^{1/2}} \frac{L}{2} \left\{ 1 + \frac{L}{4} \left( \frac{a}{ \left( \lambda_{n}^{2} - a s \right)} \right) \right\} \right]^{-1/2}.
\eeq
Plugging Eq.~\eqref{lambda} into Eq.~\eqref{An1}, we finally obtain Eq.~\eqref{A_cte_L}.
%%%%%%%%%%%%%%%%%%%%%%%%%%%%%%%%%%%%%%%%%%%%%%%%%%%%%%%%%%%%%%%%%

\bibliography{refs}
\bibliographystyle{apsrev4-1}

\end{document}